%% file: scis_paper.tex
\documentclass{SCIS2023}

\begin{document}


\ArticleType{RESEARCH PAPER}
\Year{2023}
\Month{}
\Vol{}
\No{}
\DOI{}
\ArtNo{}
\ReceiveDate{}
\ReviseDate{}
\AcceptDate{}
\OnlineDate{}

\title{Post-Layout Simulation Driven Analog Circuit Sizing}{Post-Layout Simulation Driven Analog Circuit Sizing}

\author[2,1]{Xiaohan Gao}{}
\author[1]{Haoyi Zhang}{}
\author[1]{Siyuan Ye}{}
\author[4]{Mingjie Liu}{}
\author[4]{David Z.~Pan}{}
\author[1]{Linxiao Shen}{}
\author[1, 3]{\\ Runsheng Wang}{}
\author[1, 3]{Yibo Lin}{{yibolin@pku.edu.cn}}
\author[1, 3]{Ru Huang}{}

\AuthorMark{Xiaohan Gao}

\AuthorCitation{Xiaohan Gao, Haoyi Zhang, Siyuan Ye, et al}


\address[1]{School of Integrated Circuits, Peking University, Beijing {\rm 100871}, China}
\address[2]{School of Computer Science, Peking University, Beijing {\rm 100871}, China}
\address[3]{Institute of Electronic Design Automation, Peking University, Wuxi {\rm 214000}, China}
\address[4]{Department of Electrical and Computer Engineering, University of Texas at Austin, Austin {\rm 78712}, USA}

\abstract{Post-layout simulation provides accurate guidance for analog circuit design, but post-layout performance is hard to be directly optimized at early design stages.
Prior work on analog circuit sizing often utilizes pre-layout simulation results as the optimization objective.
In this work, we propose a post-layout-simulation-driven (post-simulation-driven for short) analog circuit sizing framework that directly optimizes the post-layout simulation performance.
The framework integrates automated layout generation into the optimization loop of transistor sizing and leverages a coupled Bayesian optimization algorithm to search for the best post-simulation performance.
Experimental results demonstrate that our framework can achieve over 20\% better post-layout performance in competitive time than manual design and the method that only considers pre-layout optimization.}


\maketitle

\input{doc/intro}
\input{doc/prelim}
\input{doc/algo}
\input{doc/result}
\input{doc/conclu}
\input{doc/acknowledgement}




\input{scis_paper.bbl}




\end{document}

%% file: doc/intro.tex
\section{Introduction}
\label{sec:Introduction}

Analog circuit sizing is a critical task in analog design automation.
The sizing configuration of transistors not only dominates the performance in the schematic design but also has a high impact on the subsequent layout design.
Previous analog sizing methods based on pre-layout schematic simulation become incompatible with the increasingly sophisticated performance requirements, which drives the emerging research on analog circuit sizing aware of post-layout effects.
With the increasingly sophisticated performance requirements, analog sizing methods only considering the pre-layout simulation are no longer enough.
Thus, analog circuit sizing is evolving toward the awareness of post-layout performance.


Conventional analog design flow can be split into schematic phase and layout phase.
In the pre-layout phase, an analog designer develops a sizing configuration to meet a performance specification according to the results of the pre-layout simulation.
In the conventional design flow, layout design requires end-to-end manual participation.
Therefore, it is difficult to consider post-layout performance at the circuit sizing stage, as re-sizing the schematic design could lead to re-drawing the layout manually, which could be extremely time-consuming especially when such a procedure needs to iterate for design closure.
However, the emergence of fully automated layout generation tools, 
such as ALIGN~\cite{kunal2019align}, BAG~\cite{crossley2013bag, chang2018bag2}, and MAGICAL~\cite{xu2019magical}, 
bring new opportunities for the revolution of the existing fully manual design methodology.
As these tools are designed under the philosophy of ``no-human-in-the-loop'', they can be integrated into the optimization loops for efficient layout generation.
The insuperable barrier between the pre-layout design phase and the layout design phase 
can be broken.

With the technology node shrinking to a smaller scale and the performance requirements becoming more specialized, the discrepancy between pre-layout schematic simulation and post-layout simulation cannot be ignored.
Recent works on analog circuit sizing unveil the insufficiency of simply relying on the pre-layout simulation alone for analog circuit sizing.
Liu et al.~\cite{liu2021parasitic} propose a post-layout parasitic-aware circuit sizing method using graph neural network as a surrogate model.
Ranjan et al.~\cite{ranjan2004fast} take the post-layout parasitics effects into account with a symbolic performance model.
BagNet~\cite{hakhamaneshi2019bagnet} introduces a sizing framework that uses a deep neural network to select a sizing configuration, generate a layout, and extract parasitics for the selected sizing configuration.
AutoCkt~\cite{settaluri2020autockt} utilizes deep reinforcement learning for circuit synthesis taking post-layout parasitics into account.
These works integrate the aforementioned layout generation tools as the internal layout generators.
However, none of those works consider the post-simulation performance as the direct optimization objective.
Moreover, most works leverage neural networks as surrogate models, which usually require large amounts of data to achieve high accuracy for finding high-quality solutions \cite{bengio2017deep, chai2022circuitnet}. 
To accelerate the analog design closure, we should go one step further to optimize the post-simulation performance directly.


Despite the diverse modeling of analog circuit sizing problem, most of the sizing works treat the sizing solution synthesis as a black-box optimization problem.
Early attempts include using simulated annealing on symbolic AC models, geometric programming, and evolutionary algorithms~\cite{boyd2001optimal}.
Recent work~\cite{wang2020gcn} also explores reinforcement learning combined with graph neural networks for transferable transistor sizing.
Li et al.~\cite{li2021circuit} propose an actor-critic reinforcement learning approach to optimize post-simulation performance.
However, they do not combine pre-layout simulation with post-layout simulation to consider the sizing process.
Besides, Bayesian optimization is also adopted in analog circuit sizing for optimizing the pre-layout performance~\cite{DAC_2019, zhang2020efficient}, as it is suitable for problems with objectives expensive to compute.
Recent work considers post-layout performance with Bayesian fusion technique~\cite{bayes_fusion, transfer_bayesian_sampling}, which calculates the initial performance model and requires an additional training process.


To enable direct optimization for the post-layout performance,
we propose a post-simulation-driven analog circuit sizing framework.
We create a Bayesian optimization paradigm of two coupled Bayesian optimization models.
The specially designed framework can perform well in terms of efficiency as well as post-layout performance.

We summarize our major contributions as follows:

\begin{itemize}
    \item This work directly optimizes post-layout performance at the analog circuit sizing stage.
    \item We propose a new paradigm of coupled Bayesian optimization and apply this optimization technique to the sizing problem, to leverage both the efficiency of pre-layout simulation and the precision of the post-layout simulation.
    \item Experimental results on real-world analog circuit designs demonstrate the advantages of our framework to enhance the post-layout performance in terms of multiple metrics.
    
\end{itemize}

The rest of this paper is organized as follows.
Section~\ref{sec:Preliminary} recalls the pre-layout simulation and post-layout simulation (post-simulation for short), and introduces the basic idea of Bayesian optimization.
In the last part of Section~\ref{sec:Preliminary}, we formulate the post-simulation-driven analog circuit sizing problem.
Section~\ref{sec:Algorithm} takes a journey on the algorithm details of our proposed framework.
Section~\ref{sec:Results} discusses how we set up our experiment and demonstrates the experimental results.
Section~\ref{sec:Conclusion} summarizes the whole paper.

%% file: doc/prelim.tex
\section{Preliminaries}
\label{sec:Preliminary}

In this section, we introduce the background of our framework, including pre-layout simulation, post-layout simulation, and Bayesian optimization.
Besides, we formulate the post-simulation-driven circuit sizing problem.

\subsection{Pre-layout Simulation v.s. Post-layout Simulation}

\begin{figure}
    \centering
    \includegraphics[width=0.55\textwidth]{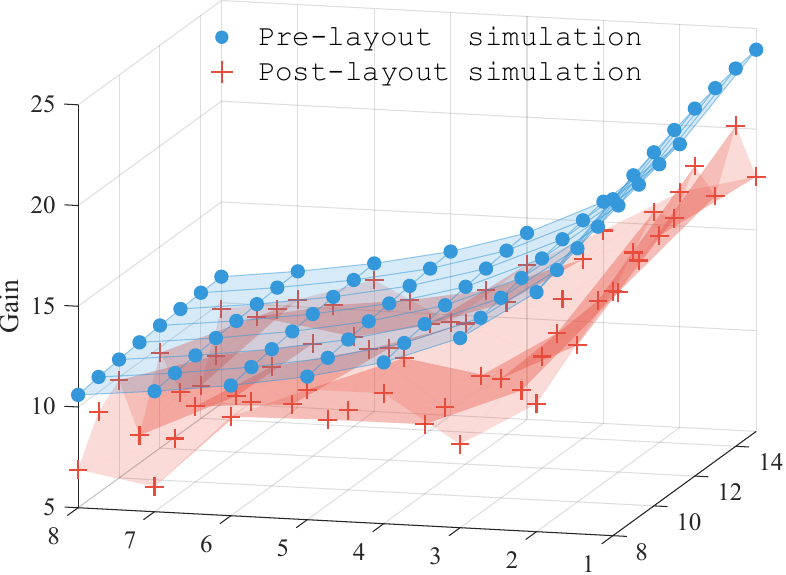}
    \caption{Pre-layout and post-layout simulation results on Gain varying with two sizing parameters.}
    \label{fig:prepost}
\end{figure}

Previous works on analog circuit sizing often utilize pre-layout simulation as a performance modeling method~\cite{DAC_2019}.
Figure~\ref{fig:prepost} illustrates an example of pre-layout simulation results and post-layout simulation results on loop gain.
We plot the results on the gain results varying with two selected sizing parameters.
The gain results of pre-layout simulation form the surface of blue dotted points and the gain results of post-layout simulation form the surface of red cross points.
As we can see, the pre-layout simulation results are generally more optimistic than the post-layout simulation.
The surface of pre-layout simulation results varies much more smoothly with the two parameters.
From a global perspective, the pre-layout simulation results turn out to have a similar trend to the post-layout simulation.
However, when diving into a local region, the pre-layout simulation cannot model the accurate post-layout performance.
Note that this figure is a direct view of how one metric varies with two parameters, while a realistic sizing scenario involves multiple metrics and a series of sizing parameters.
For the high-dimensional optimization problem, the surfaces of pre-layout simulation and post-layout simulation in the high-dimensional space will be way more different.

The results of the pre-layout simulation can neither reflect the interconnect parasitics nor all the other complex layout-dependent effects on the circuit performance, which introduces a bias that cannot be ignored.
As directly optimizing the post-layout performance is desired, switching to post-layout simulation could be a potential solution.
However, post-layout simulation suffers from layout generation overhead and heavy computation costs.
It is intuitive to discover a hybrid high-performance sizing framework combining pre-layout simulation and post-layout simulation.

\subsection{Bayesian Optimization}
\label{sec:bo}

Bayesian optimization is a widely used strategy for global black-box optimization problems, especially for those expensive-to-evaluate black-box functions \cite{bergstra2011algorithms, snoek2012practical, shahriari2015taking, zoopt}.
The essence of Bayesian optimization is rooted in surrogate model and acquisition function.
Surrogate model is supposed to approximate the objective function and quantify the uncertainty on the posterior distribution.
Acquisition function proposes sampling points and evaluates the usefulness of the sampled point for maximizing the objective function.

The most commonly used surrogate model is Gaussian process (GP)~\cite{rasmussen2003gaussian}.
A Gaussian process $\mathrm{GP(\mu_0, k)}$ is a non-parametric model described by prior mean function $\mu_0: \mathcal{X} \mapsto \mathbb{R}$ and covariance function $k: \mathcal{X} \times \chi \mapsto \mathbb{R}$.
The unknown function values $\mathbf{f}$ are modeled as a joint Gaussian distribution.
Given $\mathbf{f}$, the noisy observations $\mathbf{y}$ will be normally distributed:

\begin{equation}
\label{equ:gaussian_process}
\begin{aligned}
\mathbf{f} \; | \; & \mathbf{X} \sim \mathcal{N}(\mathbf{m}, \mathbf{K})  \\
\mathbf{y} \; | \; & \mathbf{f}, \sigma^2 \sim \mathcal{N}(\mathbf{f}, \sigma^2 \mathbf{I})  \\
\end{aligned}
\end{equation}

where the i$^{th}$ element of the mean vector $\mathbf{m}$ is $m_i\coloneqq \mu_0(\mathbf{x}_i)$ and the $(i, j)$ element of the covariance matrix $\mathbf{K}$ is $K_{i, j} \coloneqq k(\mathbf{x_i}, \mathbf{x_j})$. The distribution on $\mathbf{f}$ gives the prior distribution $p(\mathbf{f})$ for the Gaussian process.
For a newly observed data point $\mathbf{x}$, the conditional probability of random variable $f(\mathbf{x})$ on previous observations $\mathcal{D}$ is normally distributed:

\begin{equation}
\label{equ:gaussian_process_inference}
\begin{aligned}
\mu(\mathbf{x}) = & \mu_0(\mathbf{x})+\mathbf{k}(x)^T(\mathbf{K}+\sigma^2\mathbf{I})^{-1}(\mathbf{y-m})  \\
\sigma^2(\mathbf{x}) = & \kappa(\mathbf{x}, \mathbf{x})-\mathbf{k}(\mathbf{x})^T(\mathbf{K}+\sigma^2\mathbf{I})^{-1}\mathbf{k}(\mathbf{x})   \\
\end{aligned}
\end{equation}

where $\mathbf{k(x)}$ is the covariance between $\mathbf{x}$ and $\mathbf{x}_{1:n}\in \mathcal{D}$. The posterior functions will guide the selection for the next data point to be observed.

A key challenge for the acquisition function is to trade off exploitation and exploration.
Exploitation means sampling the next data point in which the prediction of the surrogate model is high, while exploration means sampling the next data point in which the surrogate model is highly uncertain.
That is, exploration encourages exploring the objective function space and exploitation encourages finding a high objective.
Common mechanisms for acquisition function include expected improvement~\cite{mockus1978application}, probability of improvement~\cite{kushner1964new}, Thompson sampling~\cite{thompson1933likelihood}, and entropy search portfolio~\cite{hennig2012entropy}.

The expected improvement (EI) is a widely applicable mechanism for acquisition function.
The expected improvement is defined as:

\begin{equation}
\label{equ:ei}
\mathrm{EI}_{y^{t}}(\mathbf{x}) \coloneqq \int_{-\infty}^{\infty} \max(y^{t}-y, 0)p_{M}(y|x) \mathrm{d}y 
\end{equation}

where $y^t$ is a designated threshold, and $M$ is some model for black-box function $f: \mathcal{X}\to \mathbb{R}^N$.
$\mathrm{EI}$ is the expectation under model $M$ that $f$ exceeds the threshold $y^t$.
The criteria of the expected improvement mechanism is to optimize $\mathrm{EI}$ to approximate $f(x)$.


\subsection{Problem Formulation}
\label{sec:problem_formulation}

The goal of our framework is to optimize the post-layout performance in the circuit sizing stage.
The measurement of analog circuit performance often involves multiple performance metrics.
Thus, the modeling of the analog circuit sizing can be seen as a multi-objective constrained optimization.
To better formulate the sizing problem, we adopt the figure-of-merit (FOM) representation.
We describe FOM formulation for analog sizing as follows:

\begin{equation}
\label{equ:fom_def}
\begin{aligned}
\max_{P} \quad & \mathcal{FOM}(P) = \sum_{i}\alpha_i f_i(P) \\
\textrm{s.t.} \quad & thres_{low}^j \leq f_j(P) \leq thres_{high}^j \\
\end{aligned}
\end{equation}

where $P$ represents the sizing configuration, $f_i(\cdot)$ and $f_j(\cdot)$ denote the metrics of post-layout performance, $\alpha_i$ is the corresponding coefficient for the i$^{th}$ metric term in the FOM, and $thres^j$ are thresholds to constrain the j$^{th}$ metric.
Each sizing configuration $P$ represents a set of assignments for all sizing parameters, including the number of fingers and finger width for each transistor.
We define the sizing parameter space as the high-dimensional space of all feasible assignments for sizing parameters, and the sizing configuration is sampled from the sizing parameter space.
The proposed framework aims to maximize $\mathcal{FOM}$ without constraint violation over sizing parameters and finally provides a good sizing solution that each metric performs well in post-layout simulation. 

%% file: doc/algo.tex
\section{Algorithm}
\label{sec:Algorithm}

In this section, we introduce our analog circuit sizing framework driven by post-layout simulation.
The framework builds up a compact closure considering the post-layout simulation at early sizing stage.
We take advantage of automatic layout generation to integrate the post-layout simulation into the closure.
For efficiency considerations, we utilize pre-layout simulation to compensate for the time-consuming post-simulation loop.
We design a coupled Bayesian optimization that makes use of pre-layout simulation results to explore the parameter space and apply post-layout simulation results to guide the best FOM exploiting.
As analog circuit sizing involves discrete parameters such as the number of fingers for a transistor, we consider the Bayesian estimator for mixed-variable optimization. 
We will detail the proposed framework in the following subsections. 

\subsection{Overview of Our Framework}
\label{sec:overview}

{
\begin{figure}
    \centering
    \includegraphics[width=\textwidth]{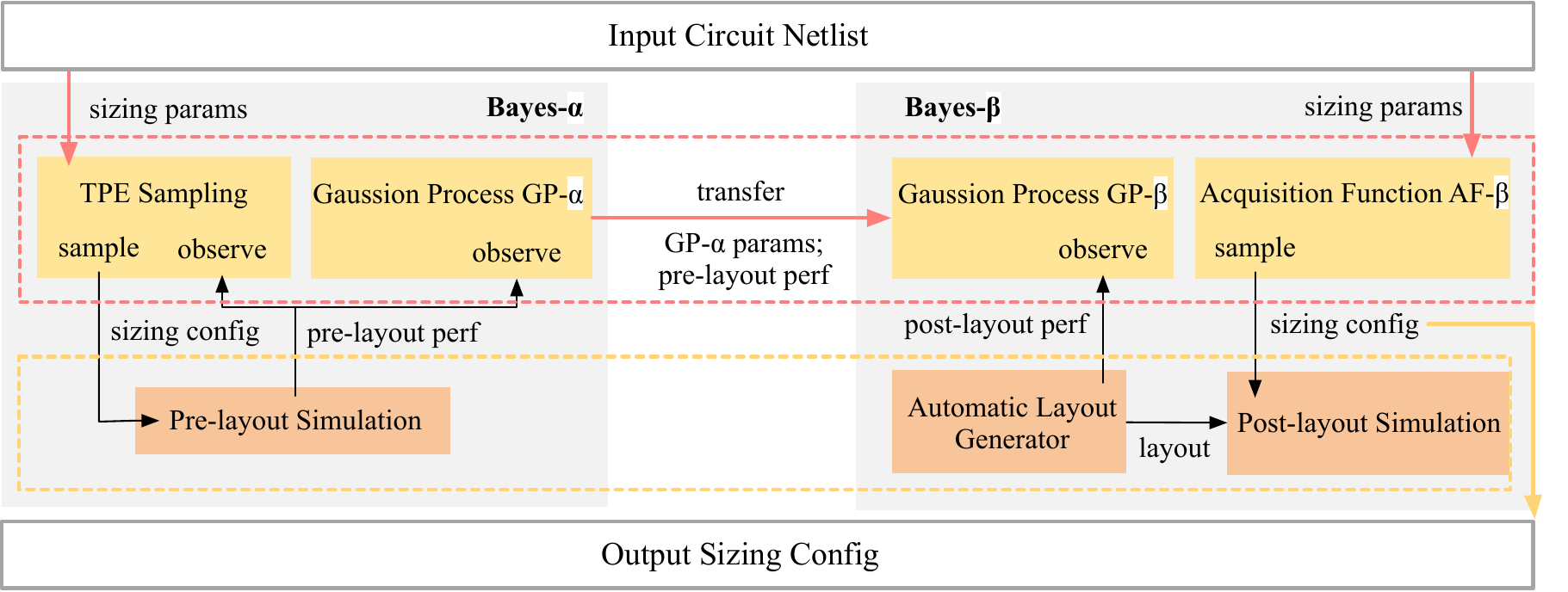}
    \caption{The overall flow of our proposed framework.}
    \label{fig:flow}
\end{figure}
}

Figure~\ref{fig:flow} shows the workflow of our framework.
Different from previous analog sizing work, we propose a novel coupled Bayesian optimization paradigm.
Our framework combines two models named \texttt{Bayes}-$\alpha$ and \texttt{Bayes}-$\beta$.
\texttt{Bayes}-$\alpha$ consists of a tree-structured Parzen estimator and a Gaussian process.
\texttt{Bayes}-$\alpha$ applies Bayesian optimization on the black-box function of pre-layout simulation.
\texttt{Bayes}-$\beta$ attempts to optimize the black-box function of a more complex procedure with an automatic layout generator and post-layout simulation in it.
The intrinsics of the two black-box functions are the pre-layout performance and post-layout performance correspondingly.

Our framework combines the two GP-based Bayesian optimization parts of \texttt{Bayes}-$\alpha$ and \texttt{Bayes}-$\beta$ in a way we call coupled Bayesian optimization.
We propose the concept of coupled Bayesian optimization:

\begin{definition}
\label{def:cbs}
    A group of Bayesian optimization models is said to be coupled if: 
    
    1) they share the same form of acquisition functions, and
    
    2) their surrogate models can be seen as generating from the same rendering process with different values of hyperparameters.
\end{definition}

To be more specific, in our framework, \texttt{Bayes}-$\alpha$ and \texttt{Bayes}-$\beta$ are coupled, 
as we can transfer the belief that surrogate model $\mathrm{GP}^\alpha$ gains from the pre-layout performance black-box function to the domain of surrogate model $\mathrm{GP}^\beta$.
That is, we can make use of the observed FOM values in pre-layout simulation to benefit the sampling in Bayesian optimization for post-layout performance.
We will discuss more about how to realize this powerful property in Section~\ref{sec:mcmc}.

As we mentioned in Section~\ref{sec:bo}, a well-designed acquisition function is supposed to trade off exploration and exploitation in a balanced way.
For two coupled Bayesian optimization models, there is a different way to trade off exploration and exploitation.
Our framework makes \texttt{Bayes}-$\alpha$ explore the sizing parameter space and \texttt{Bayes}-$\beta$ focus on exploiting the higher FOM values.
We will tackle sampling from the mixed parameter space for \texttt{Bayes}-$\alpha$ in Section~\ref{sec:discrete}.
Furthermore, we show how we do efficient exploration with a tree-structured Parzen estimator for \texttt{Bayes}-$\alpha$ in Section~\ref{sec:acqf}.

As the coupled Bayesian optimization is an iterative process, our framework conducts iterations with the two Gaussian processes $\mathrm{GP}^\alpha$ and $\mathrm{GP}^\beta$.
Frequent calls of post-layout simulation can be extremely time-consuming, and therefore we set a variable $interval$ to control how many iterations we run one single post-layout simulation loop.
Before sampling from \texttt{Bayes}-$\beta$, we update \texttt{Bayes}-$\beta$ with \texttt{Bayes}-$\alpha$.

\subsection{Transfer Surrogate Model}
\label{sec:mcmc}

As the computation cost of automatic layout generation and post-layout simulation is far greater than pre-layout simulation, our framework explores how to make less observation in \texttt{Bayes}-$\beta$ but manages to approximate the best FOM of post-layout simulation. 
To benefit the \texttt{Bayes}-$\beta$ from the richer sampling points but not accurate simulation results in \texttt{Bayes}-$\alpha$, we explore the transfer of the surrogate model in \texttt{Bayes}-$\alpha$ to the one in \texttt{Bayes}-$\beta$.

\texttt{Bayes}-$\alpha$ and \texttt{Bayes}-$\beta$ have their own GP-based surrogate model, and the Gaussian process $\mathrm{GP}^\alpha$ of \texttt{Bayes}-$\alpha$ gains the belief about the black-box function $f^{\alpha}$.
To clarify, we use the notation $f^{\alpha}$ for the black-box function of pre-layout simulation, and $f^{\beta}$ for the black-box function of post-layout simulation.
A naive strategy is to regard the Gaussian process $\mathrm{GP}^\alpha$ and $\mathrm{GP}^\beta$ as two independent processes, but the strategy will not work due to the lack of observed data points for $\mathrm{GP}^\beta$.
We propose a strategy to gain belief about black-box $f^\beta$ from both $\mathrm{GP}^\beta$ and $\mathrm{GP}^\alpha$.

Our strategy is inspired by the re-parameterization trick~\cite{kingma2013auto, rezende2014stochastic}.
Note that for an arbitrary parameter configuration $\mathbf{x}$, the value $f^\alpha(\mathbf{x})$ and $f^{\beta}(\mathbf{x})$ are strongly correlated in some way unknown.
We assume that they are jointly distributed as:

\begin{equation}
\label{equ:joint_distributed}
\begin{bmatrix} f^\alpha(\mathbf{x}) \\ f^\beta(\mathbf{x}) \end{bmatrix}
 \sim
 \mathcal{N}
 \begin{pmatrix}
  \begin{bmatrix}
   \mu^\alpha \\ \mu^\beta
  \end{bmatrix},
  &
  \begin{bmatrix}
   \mathbf{K^\alpha} & \mathbf{K}^{\alpha, \beta} \\
   \mathbf{K}^{\beta, \alpha} & \mathbf{K}^\beta
  \end{bmatrix}
 \end{pmatrix}
\end{equation}

where $\mu^\alpha$, $\mu^\beta$ are mean vectors, and $\mathbf{K}^\alpha$, $\mathbf{K}^\beta$ are Gaussian process kernels, defined in Equation~\ref{equ:gaussian_process_inference}.
$\mathbf{K}^{\alpha, \beta}$ and $\mathbf{K}^{\beta, \alpha}$ are covariance matrices between $f^\alpha$ and $f^\beta$.

Consider GP$^\alpha$ and GP$^\beta$ as two models generated by the same sampler from two different sets of hyperparameters.
We adapt the re-parameterization trick to formulate the posterior distribution~\cite{kingma2013auto}.
For a new observation of parameter configuration $(\mathbf{x}, f^\alpha(\mathbf{x}), f^\beta(\mathbf{x}))$, we first update $\mathbf{K}$ with general Bayesian optimization strategy, and then update $\mathbf{K}^\beta$ for GP$^\beta$:
\begin{equation}
\label{equ:K_update}
\begin{aligned}
K^\beta = K^\beta - K^{\beta,\alpha} (K^{\alpha})^{-1} K^{\alpha, \beta}
\end{aligned}
\end{equation}


\subsection{Tackle Mixed-variable Optimization}
\label{sec:discrete}


Traditional Bayesian optimization methods usually deal with black-box functions with continuous variables. 
Analog circuit sizing faces optimizing continuous variables and discrete variables simultaneously.
The Bayesian optimization does not naturally support mixed-variable optimization.
To enhance the optimization process, we design our acquisition function to sample from both continuous and discrete variables, based on the tree-structured Parzen estimator approach.

The tree-structured Parzen estimator (TPE) approach derives from the expected improvement optimization scheme~\cite{bergstra2011algorithms}.
As we mentioned before, the Gaussian-process-based method models $p(y|\mathbf{x})$, and furthermore TPE models both $p(y|\mathbf{x})$ and the distribution of objective function values $p(y)$.
We inherit the replacing strategy from the tree-structured Parzen estimator that models $p(y|\mathbf{x})$ by replacing the prior distributions on sizing parameter space with non-parametric densities.
The most used parameters like the number of fingers and finger width can be described with uniform variables and quantized uniform variables.
Under the replacing strategy, for example, we can describe a parameter of finger width as a truncated Gaussian mixture.
Therefore, we can produce a variety of densities over the sizing parameter space $\mathcal{X}$ and we define two densities for sizing parameter space:

\begin{equation}
\label{equ:lebesgue}
\begin{aligned}
\ell(\mathbf{x})  = & \frac{\partial x\ast \mathbf{P^\ell}}{\partial \mu_{\mathcal{X}}},  x\in \{x^i| f(x^i)< y^*\}  \\
g(\mathbf{x}) = & \frac{\partial x\ast \mathbf{P}^g}{\partial \mu_{\mathcal{X}}}, x\in \{x^i | f(x^i)\geq y^*\} \\
\end{aligned}
\end{equation}

where $y^*$ is the best FOM value observed yet, $\ell(x)$ represents the density induced from the observations $\{x^i\}\in \mathcal{X}$ whose prediction is less than $y^*$, $g(\mathbf{x})$ represents the density induced from other observed $x^i$.
$x\ast \mathbf{P^\ell}$ is the probability measure on measurable space $(\mathcal{X}, \mathcal{A}^\ell)$, $\mathcal{A}^\ell\coloneqq \{x^i|f(x^i)\leq y^*\}$ and $\mu_{\mathcal{X}}$ is a reference measure.
Equation~\ref{equ:lebesgue} defines how we get the two densities $\ell(\mathbf{x})$ and $g(\mathbf{x})$ by Radon-Nikodym derivative.
We adopt the formulation from~\cite{srivastava1973estimation} to numerically calculate the $\ell(\mathbf{x})$ and $g(\mathbf{x})$. Here we gives how we calculate $\ell(\mathbf{x})$ as example:

\begin{equation}
\label{equ:parzen}
\begin{aligned}
\ell(\mathbf{x})  = & \frac{1}{n^{\ell}h(n^{\ell})}\sum_{x^i} K(\frac{\mathbf{x}-x^i}{h(n^\ell)}), \quad x^i \in \mathcal{A}^\ell \\
\end{aligned}
\end{equation}

where $n^{\ell}$ stands for the size of set $\mathcal{A}^\ell$, $K(\cdot)$ is a real-valued Borel function and $h(\cdot)$ is a sequence of positive real numbers, and they satisfy the conditions described in~\cite{srivastava1973estimation}.
As we can calculate the Equation~\ref{equ:parzen} for both continuous and discrete variables, the density definition is applicable to all parameter spaces met in the analog circuit sizing problem. 
We will continue the subsequent deduction for acquisition function design in the next subsection. 

\subsection{Control the Acquisition Function}
\label{sec:acqf}

As we expect \texttt{Bayes}-$\alpha$ to explore more parameter space, a controlling strategy for TPE is desired.
We adapt the tree-structured Parzen estimator~\cite{bergstra2011algorithms} to control whether to sample aggressively.
Sampling aggressively implies more exploitation and sampling less aggressively indicates more exploration.
Following the Equation~\ref{equ:lebesgue} and ~\ref{equ:parzen}, 
Then we can define the $p(\mathbf{x}|y)$ on sizing parameter space using the pre-defined $\ell(\mathbf{x})$ and $g(\mathbf{x})$:

\begin{equation}
\label{equ:density_tpe}
  p(y|\mathbf{x}) =
    \begin{cases}
      \ell(\mathbf{x}) & y < y^* \\
      g(\mathbf{x}) & y \geq y^*
    \end{cases}  
\end{equation}

Under the assumption of Equation~\ref{equ:density_tpe}, we can design acquisition functions for our framework to optimize the expected improvement introduced in Equation~\ref{equ:ei}.
By a simple application of the Bayesian theorem, we derive:

\begin{equation}
\label{equ:ei_cond}
\mathrm{EI}_{y^*}(\mathbf{x}) = \int_{-\infty}^{y^*}(y^*-y)\frac{p(\mathbf{x}|y)p(y)}{p(\mathbf{x})}\mathrm{d} y
\end{equation}

here we take $y^*$ for the threshold $y^t$.
As the marginalization property indicates, we have:

\begin{equation}
\label{equ:marginal}
p(\mathbf{x}) = \int_{\mathbb{R}}p(\mathbf{x}|y)p(y)\mathrm{d} y
\end{equation}

from which we deduce that:

\begin{equation}
\label{equ:ei_density}
\begin{aligned}
\int_{-\infty}^{y^*}(y^*-y)p(\mathbf{x}|y)p(y)\mathrm{d}y=\ell(\mathbf{x})\int_{-\infty}^{y^*}(y^*-y)p(y)\mathrm{d}y \\
= p(y<y^*)y^*\ell(\mathbf{x})-\ell(\mathbf{x})\int_{-\infty}^{y^*}p(y)\mathrm{d}y
\end{aligned}
\end{equation}

Therefore, we get the proportional formulation:

\begin{equation}
\label{equ:ei_re}
\begin{aligned}
\mathrm{EI}_{y^*}(\mathbf{x}) = \frac{p(y<y^*)y^*\ell(\mathbf{x})-\ell(\mathbf{x})\int_{-\infty}^{y^*}p(y)dy}{p(y<y^*)\ell(\mathbf{x})+p(y\geq y^*)g(\mathbf{x})}
\end{aligned}
\end{equation}

\begin{equation}
\label{equ:ei_prop}
\begin{aligned}
\mathrm{EI}_{y^*}(\mathbf{x}) \propto  (p(y<y^*)+\frac{g(\mathbf{x})}{\ell(\mathbf{x})p(y\geq y^*)})^{-1}
\end{aligned}
\end{equation}

From Equation~\ref{equ:ei_prop}, we know that our proposed acquisition function will be more likely to sample the next sizing parameter configuration $\mathbf{x}$ from parameter space $\mathcal{X}$ with higher $\frac{g(\mathbf{x})}{\ell(\mathbf{x})}$.

As we discussed before, we expect the Bayesian optimization \texttt{Bayes}-$\alpha$ to explore more about the parameter space.
We assign a less aggressive threshold, that is, a small $y^*$ for \texttt{Bayes}-$\alpha$, to encourage exploring more parameter space.
 

%% file: doc/result.tex
\section{Experimental Results}
\label{sec:Results}

In this section, we demonstrate the efficiency of our proposed framework.
We conduct experiments on two real-world analog circuits, including an inverter-based operational transconductance amplifier (OTA) and a low-dropout regulator (LDO).
The two circuits are designed under TSMC 40nm technology.
We conduct our experiments on a CentOS workstation with an Intel Xeon Gold 5218R CPU and 128GB memory.
We adopt the MAGICAL~\cite{xu2019magical, zhu2020effective, chen2020toward} as our automatic layout generator.
All the pre-layout simulation results and post-layout simulation results are generated from Cadence Virtuoso and Mentor Graphics Calibre.

{
We implement a baseline considering pre-layout simulation (denoted as \texttt{Plain}) and another baseline considering both pre-layout simulation and post-layout simulation (denoted as \texttt{BMF}).
The former adopts the basic Bayesian optimization method from~\cite{liu2021parasitic}.
The latter applies the Bayesian model fusion technique~\cite{bayes_fusion} to train a post-layout performance Bayesian model and then uses the Bayesian model to find a sizing configuration.
We set the number of post-layout training samples the same as the post-layout simulation iterations.
The number of iterations in the follow-up Bayesian optimization also takes the same value.
We list the specific values in the following Section~\ref{sec:results_ota} and~\ref{sec:results_ldo}.

We estimate the runtime by program execution time other than CPU time.
We demonstrate the execution pattern of our optimization flow with Figure~\ref{fig:timebar}.
\texttt{Bayes}-$\alpha$ and \texttt{Bayes}-$\beta$ run asynchronously in most of the time.
\texttt{Bayes}-$\beta$ messages \texttt{Bayes}-$\alpha$ a sampled sizing configuration at the beginning of a \texttt{Bayes}-$\beta$ iteration.
\texttt{Bayes}-$\alpha$ messages \texttt{Bayes}-$\beta$ the parameters of Gaussian process GP$^\alpha$ and the pre-layout simulation result of the sampled sizing configuration.
As for \texttt{BMF}~\cite{bayes_fusion}, we first run pre-layout simulations and select a part of sizing configurations with top performance for the training, which is a sequential execution pattern.
}

{
\begin{figure}
    \centering
    \includegraphics[width=0.91\textwidth]{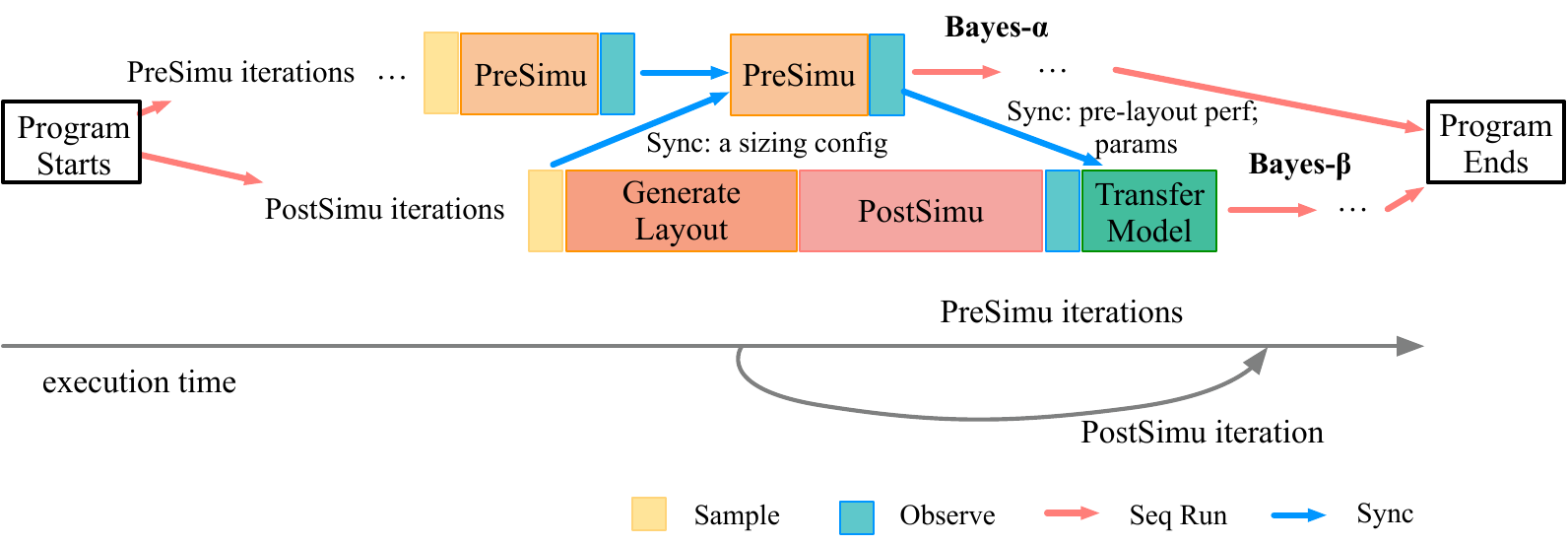}
    \caption{The asynchronous execution pattern of our optimization flow with pre-layout simulation and post-layout simulation overlapping on the execution time.}
    \label{fig:timebar}
\end{figure}
}

\subsection{Inverter-based OTA}
\label{sec:results_ota}

\begin{figure}
    \centering
    \includegraphics[width=0.98\textwidth]{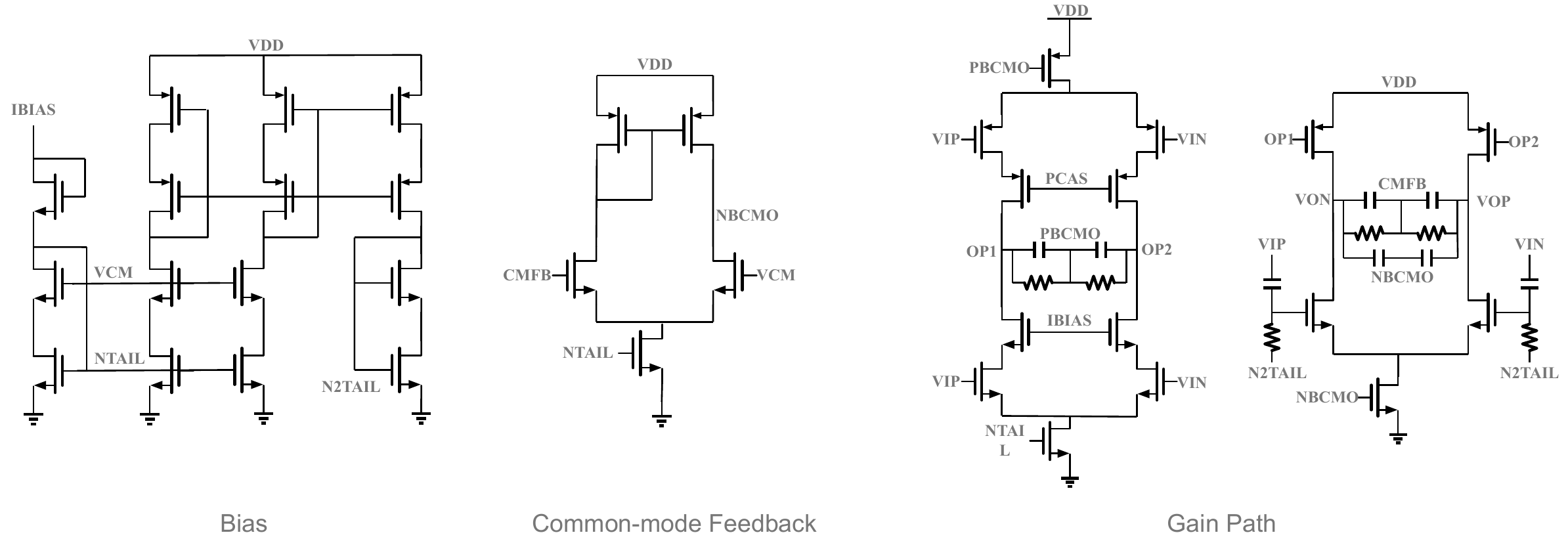}
    \caption{The schematic of the inverter-based operational transconductance amplifier.}
    \label{fig:schematic_ffc_ota}
\end{figure}

The schematic of the OTA is shown in Figure~\ref{fig:schematic_ffc_ota}. 
This circuit is reproduced from~\cite{xu2019magical}, which is a taped-out case. 

We perform symmetry detection for this OTA circuit~\cite{liu2020s} and reduce the design parameters to 20 for the sizing process.
To meet a reasonable design requirement, we construct the specific form of FOM for this circuit:

\begin{equation}
\label{equ:fom_ota}
\begin{aligned}
\max_{P} \quad & \alpha_{1}Gain + \alpha_2\lg UGB  \\
\textrm{s.t.} \quad & thres_{low} \leq PM \leq thres_{high} \\
\end{aligned}
\end{equation}


where $Gain$ denotes the close loop gain, $UGB$ stands for the unity gain bandwidth, and $PM$ represents the phase margin. 
As $Gain$ involves logarithmic calculation, we add a logarithmic form for the other term $UGB$ to normalize the optimization objective. 
In this experimental setup, we consider $Gain$ and $UGB$ to be equally important, and thus we set their coefficient $\alpha_1$ and $\alpha_2$ to $1.0$ respectively.
To represent the constraint on phase margin, we introduce a penalty term to the optimization objective.
To guarantee the circuit functionality, the phase margin is supposed to range from 45 to 80, and therefore we set $thres_{low} = 45$ and $thres_{high} = 80$.
The penalty term is defined as:

\begin{equation}
\label{equ:penalty_pm}
penalty = \max(thres_{high}, PM) - \min(thres_{low}, PM)
\end{equation}

{
\begin{table}[tb]
    \centering
    \caption{\textsc{The FOM results and post-simulation performance of the inverted-based OTA.}}
    \label{tab:fom_ota}
    \resizebox{.85\textwidth}{!}{
    \begin{tabular}{ccccccc}
    \toprule
    Performance & \texttt{Plain}~\cite{liu2021parasitic} & \texttt{BMF}-16~\cite{bayes_fusion} & \texttt{BMF}-32~\cite{bayes_fusion} & \texttt{Manual}~\cite{xu2019magical} & \texttt{Ours-10} & \texttt{Ours-5} \\
    \midrule
    Gain (dB) & 17.09 & 17.19 & 20.34 & 14.15 & 19.89 & \textbf{20.52} \\
    UGB (MHz) & 7.11 & 7.35 & 8.04 & 5.11 & 8.54 & \textbf{9.03} \\
    PM ($^{\circ}$) & 45$<, <$80 & 45$<, <$80 & 45$<, <$80 & 45$<, <$80 & 45$<, <$80 & 45$<,<$80 \\
    FOM & 32.86 & 33.00 & 36.24 & 29.59 & 35.85 & \textbf{36.53} \\
    \midrule
    \#PreSimu & 160 & 160 & 160 & - & 160 & 160 \\
    \#PostSimu & - & 16 & 32 & - & 16 & 32 \\
    Runtime & 87min2s & 140min5s & 198min10s & - & 95min13s & 104min27s \\
    \bottomrule
    \end{tabular}
    }
\end{table}
}

\texttt{Plain} is the basic Bayesian optimization method on pre-layout simulation results~\cite{liu2021parasitic}.
\texttt{BMF}-16 and \texttt{BMF}-32 is the Bayesian model fusion which employs 16 and 32 post-layout training samples respectively.
The iterations on the predicted Bayesian model also take 16 and 32 respectively.
\texttt{Manual} is an open-source taped-out case from MAGICAL~\cite{xu2019magical}.
In Table~\ref{tab:fom_ota}, \texttt{Ours-10} and \texttt{Ours-5} are our methods of coupled Bayesian optimization.
The \texttt{x} of \texttt{Ours-x} represents the interval that we run post-layout simulation once every \texttt{x} runs of the pre-layout simulation.
We restrict the maximum number of iterations to 160 and initialize the optimization process with random starting points.
By iteration, we mean one single loop of sampling a sizing configuration, measuring by simulation, and updating our model.
As we know from Figure~\ref{fig:timebar}, the runtime of post-layout iteration could overlap with the runtime of several pre-layout iterations. 
We estimate the time of pre-layout simulation (25$\pm$5s), the time of layout generation (70s$\pm$50s), the time of post-layout simulation (80$\pm$15s), the time of sample\&observe ($\leq$5s), and the time of transferring model ($\leq$5s) for one iteration.
We set the timeout as 240s for the layout generator.

Table~\ref{tab:fom_ota} summarizes the post-layout performance of the inverter-based OTA circuit.
We present the performance with gain, unity gain bandwidth, and phase margin metrics.
\#PreSimu and \#PostSimu represent the number of runs of pre-layout simulation and post-layout simulation.
Both \texttt{Ours-10} and \texttt{Ours-5} outperform \texttt{Plain} and manually sized design.
\texttt{Ours-10} outperforms \texttt{BMF-16} and \texttt{Ours-5} outperforms \texttt{BMF-32}.
Our framework satisfies the constraint on phase margin.
\texttt{Ours-5} achieves a $45\%$ improvement in gain, a $76\%$ improvement in unity gain bandwidth, compared with manual sizing configuration, and a $20\%$ improvement in gain, a $27\%$ improvement in unity gain bandwidth, compared with \texttt{Plain} with Bayesian optimization on pre-layout simulation.

\subsection{Low-dropout Regulator}
\label{sec:results_ldo}

\begin{figure}
    \centering
    \includegraphics[width=0.43\textwidth]{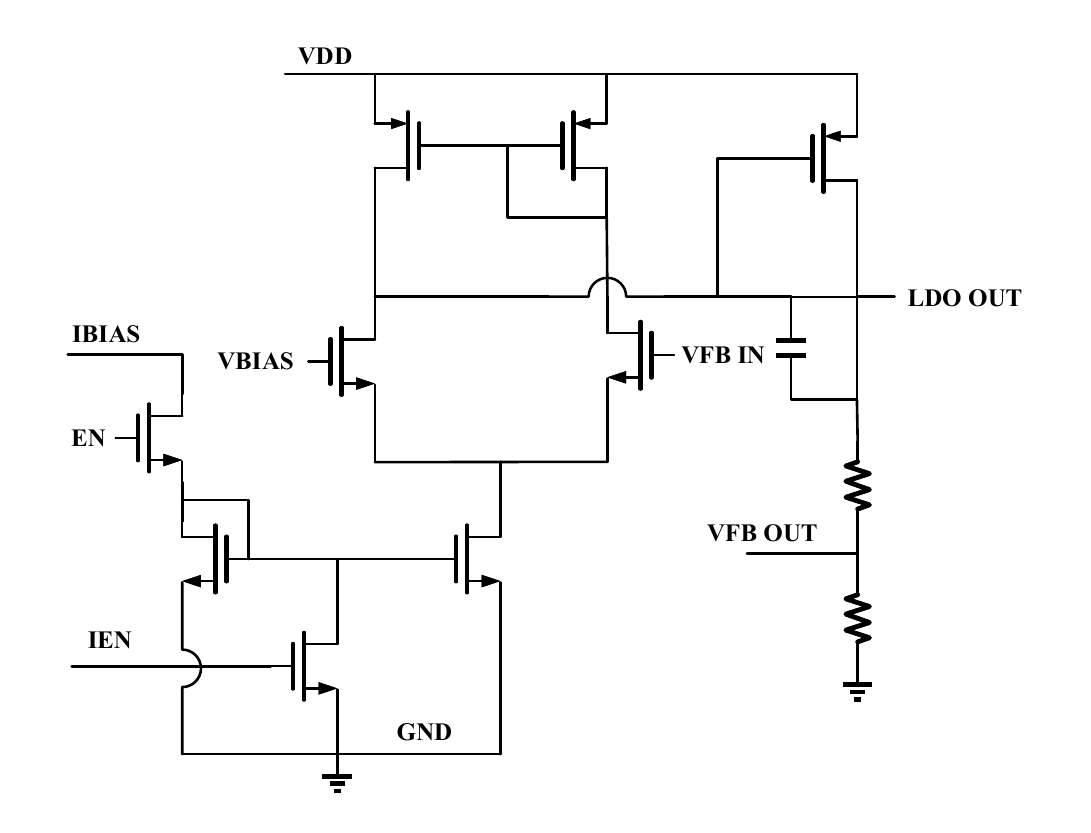}
    \caption{The schematic of the low-dropout regulator.}
    \label{fig:schematic_ldo}
\end{figure}

The schematic of the low-dropout regulator is shown in Figure~\ref{fig:schematic_ldo}. 
This circuit is designed by an experienced designer.
We use 12 design parameters for sizing this circuit. 
The specific form of FOM for this LDO circuit is defined as:

\begin{equation}
\label{equ:fom_ldo}
\begin{aligned}
\max_{P} \quad & \alpha_{1}Gain + \alpha_2 V_{OU} \\
\textrm{s.t.} \quad & thres_{low} \leq PM \leq thres_{high} \\
\end{aligned}
\end{equation}

where $Gain$ denotes the close-loop gain, $PM$ denotes the phase margin, $V_{OU}$ stands for overshot up voltage.
As the design is supposed to decrease the overshot up voltage, we assign a negative coefficient for the term. 
To be more specific, we take $\alpha_1 = 0.1$ and $\alpha_2=-10.0$ so that the two terms are of the same order of magnitude.
The penalty for phase margin constraint follows the formulation shown in Equation~\ref{equ:penalty_pm}, in which $thres_{low}$ takes $60$ and $thres_{high}$ takes $90$.

{
\begin{table}[tb]
    \centering
    \caption{\textsc{The FOM results and post-simulation performance of the Low-dropout Regulator.}}
    \label{tab:result_ldo}
    \resizebox{.85\textwidth}{!}{
    \begin{tabular}{ccccccc}
    \toprule
    Performance & \texttt{Plain}~\cite{liu2021parasitic} & \texttt{BMF}-30~\cite{bayes_fusion} & \texttt{BMF}-60~\cite{bayes_fusion} & \texttt{Manual} & \texttt{Ours-4} & \texttt{Ours-2} \\
    \midrule
    Gain (dB) & 70.16 & 72.21 & 72.71 & 73.22 & \textbf{73.45} & 72.15 \\
    OU (V) & 1.40 & 0.73 & 0.36 & 1.43 & 0.47 & \textbf{0.28} \\
    PM ($^{\circ}$) & 60$<,<$90 & 60$<,<$90 & 60$<,<$90 & 60$<,<$90 & 60$<,<$90 & 60$<,<$90 \\
    FOM & -6.98 & -0.08 & 3.63 & -6.98 & 2.65 & \textbf{4.42} \\
    \midrule
    \#PreSimu & 120 & 120 & 120 & - & 120 & 120 \\
    \#PostSimu & - & 30 & 60 & - & 30 & 60 \\
    Runtime & 54min15s & 99min2s & 143min52s & - & 96min30s & 105min50s \\
    \bottomrule
    \end{tabular}
    }
\end{table}
}

Table~\ref{tab:result_ldo} shows the results of post-layout performance for the low-dropout regulator circuit.
Here we demonstrate \texttt{Ours-4} and \texttt{Ours-2} for this case, which run one post-layout simulation per 4 and 2 iterations respectively.
We choose smaller intervals $4$ and $2$ for this case because there is a large gap between the distributions of the pre-layout simulation and the post-layout simulation of the low-dropout regulator.
Similar to the first case, we compare our framework with the manual sizing configuration \texttt{Manual} and the baseline method \texttt{Plain}, \texttt{BMF}-30, \texttt{BMF}-60.
Here \texttt{BMF}-30 and \texttt{BMF}-60 take 30 and 60 post-layout training samples respectively.
We estimate the time of pre-layout simulation (20$\pm$5s), the time of layout generation (20$\pm$15s), the time of post-layout simulation (65$\pm$15s), the time of sample\&observe ($\leq$5s), and the time of transferring model ($\leq$5s) for one iteration.
As shown in the Table~\ref{tab:result_ldo}, both \texttt{Ours-4} and \texttt{Ours-2} achieve better performance in multiple metrics.
\texttt{Ours-4} achieves the best value of gain.
\texttt{Ours-2} obtains the best value of overshot up voltage.
Taken together, our framework outperforms the manual sizing configuration and basic Bayesian optimization in FOM values within a reasonable runtime.

Figure~\ref{fig:ablation} shows the performance of our method with respect to different \#PreSimu and \#PostSimu settings.
Figure~\ref{fig:a} shows the best FOM results with different pre-layout iterations when fixing \#PostSimu=60, and Figure~\ref{fig:b} shows the results with different post-layout iterations when fixing \#PreSimu=120.
Both pre-layout and post-layout iterations contribute to the post-layout performance.
Figure~\ref{fig:c} shows the performance of \texttt{Plain} with different pre-layout iterations when their runtime is close to \texttt{Ours}-2.
We can see that barely with pre-layout simulation, the post-layout performance saturates quickly even given more simulations. 
Figure~\ref{fig:d} compares the performance over post-layout iterations between \#PreSimu=0 and \#PreSimu=120.
Note that the blue line shares the same data with Figure~\ref{fig:b}.
We can see that combining post-layout simulation with pre-layout simulation can significantly improve the efficiency of searching for better sizing configuration. 
With these experiments, we conclude that incorporating pre-layout simulation and post-layout simulation can improve both the quality of performance and efficiency. Meanwhile, our proposed algorithm can synergistically balance amounts of pre-layout simulations and post-layout simulations in an asynchronous pattern. 

{
\begin{figure}
    \centering
    \subfloat[Different \#PreSimu when \#PostSimu=60]{
    \label{fig:a}
    \includegraphics[height = 0.3\textwidth]{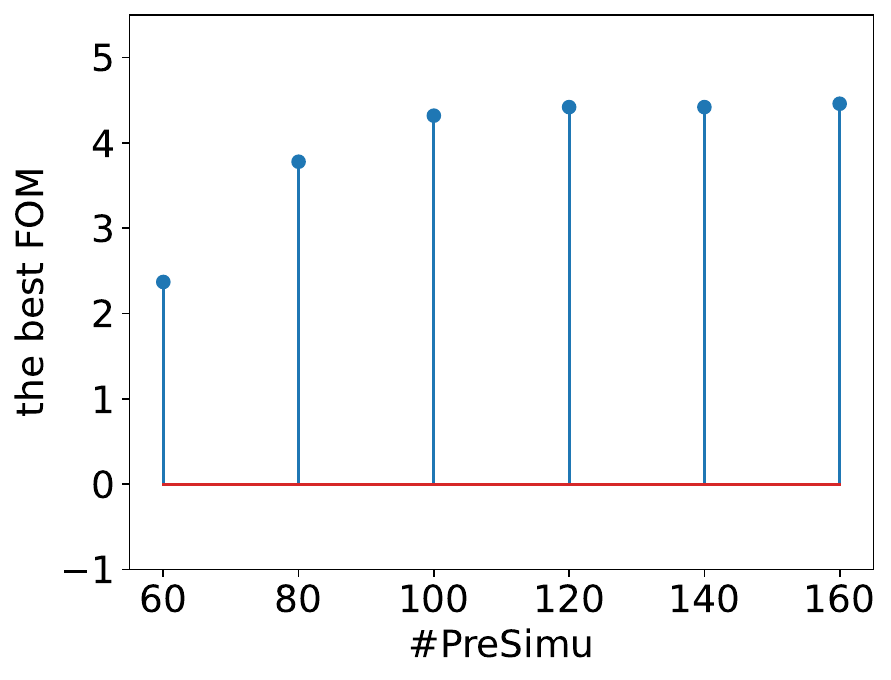}
    }
    \hspace{.15in}
    \subfloat[Different \#PostSimu when \#PreSimu=120]{
    \label{fig:b}
    \includegraphics[height = 0.3\textwidth]{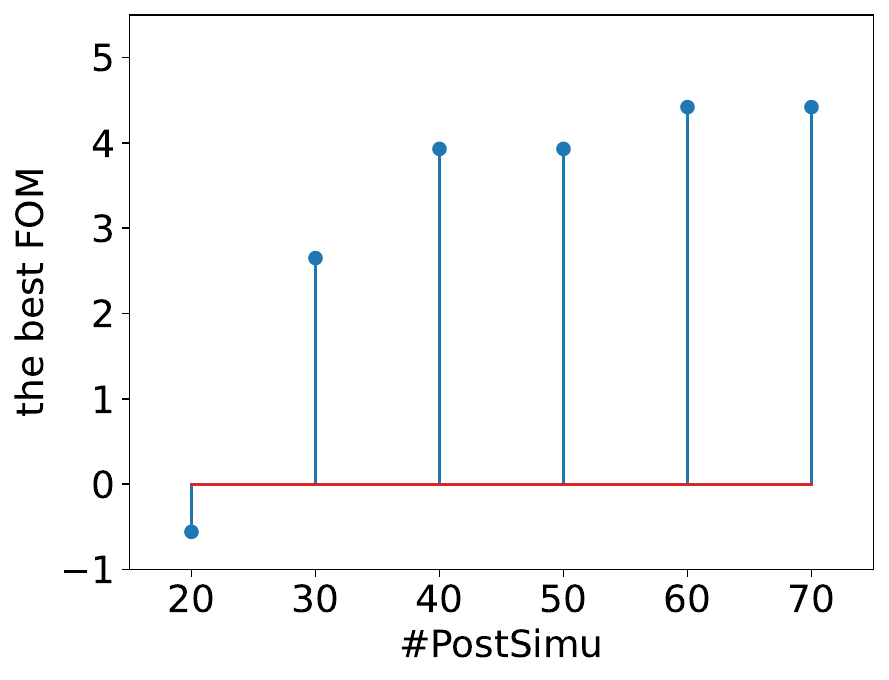}
    }

    \subfloat[Different \#PreSimu when \#PostSimu=0]{
    \label{fig:c}
    \includegraphics[height = 0.3\textwidth]{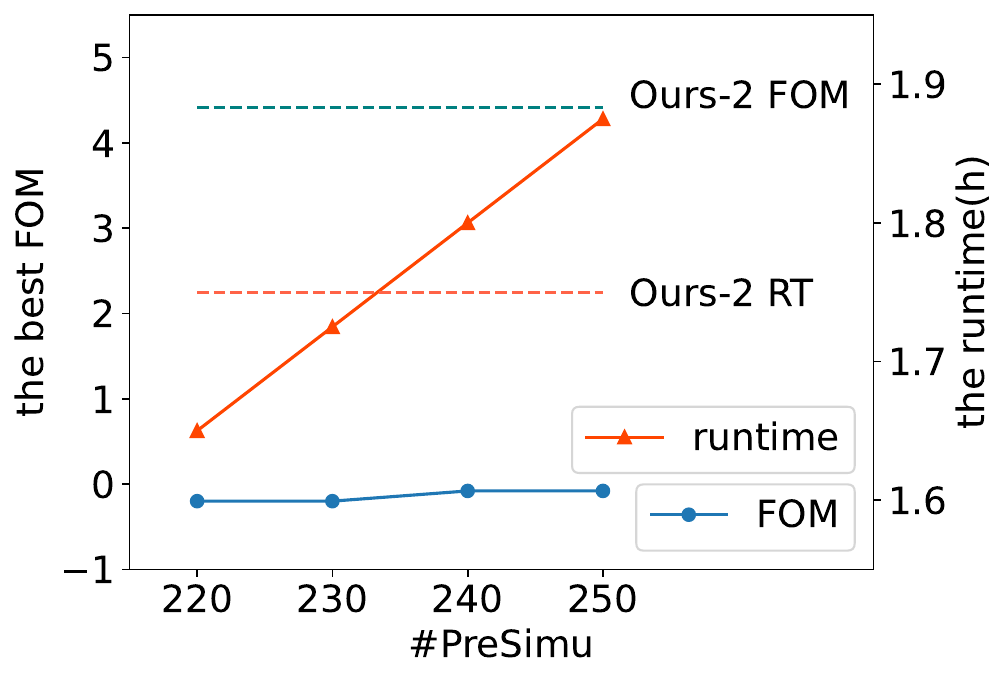}
    }
    \hspace{.15in}
    \subfloat[Comparing \#PreSimu=120 and \#PreSimu=0]{
    \label{fig:d}
    \includegraphics[height = 0.3\textwidth]{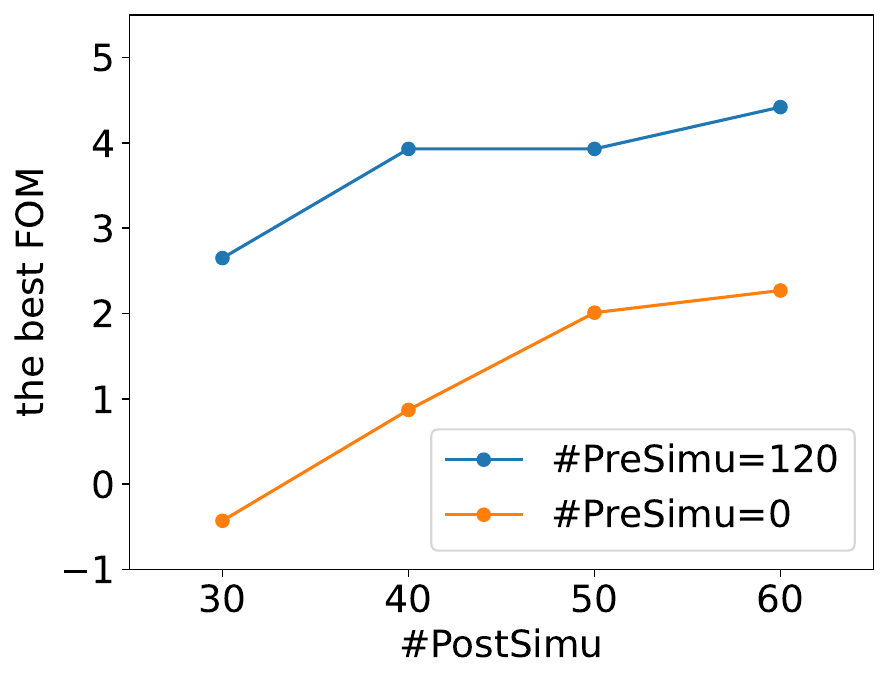}
    }
    \caption{
      The best FOM results of different \#PreSimu and \#PostSimu settings.
    }
    \label{fig:ablation}
\end{figure}
}


%% file: doc/conclu.tex
\section{Conclusion}
\label{sec:Conclusion}


In this paper, we propose a post-layout-simulation-driven analog circuit sizing framework.
By integrating the automatic layout generator and post-layout simulation into the circuit sizing loop, we develop a method that directly optimizes the post-layout performance.
Our framework leverages a coupled Bayesian optimization technique, to represent a group of strongly correlated Bayesian optimization models satisfying special properties.
The coupled Bayesian optimization helps our framework trade-off exploration and exploitation.
Our framework takes advantage of the efficiency of pre-layout simulation to explore more sizing parameter space and the post-layout simulation to exploit parameter configuration which is likely to induce a high FOM value.
Compared with the Bayesian optimization on pre-layout simulation alone and the manual sizing results, our framework is more promising for generating sizing configuration with better post-layout performance.
In future work, we can further enhance the efficiency of our proposed framework.

%% file: doc/acknowledgement.tex
\section{Acknowledge}

We would like to thank Kenuo Xu for figure drawing Figure~\ref{fig:prepost}.
This project is supported in part by 
the National Science Foundation of China (Grant No. 62141404 and No. 62034007) and 111 Project (B18001).

%% file: scis_paper.bbl

%% file: scis_paper.bbl
\begin{thebibliography}{10}
\providecommand{\url}[1]{#1}
\csname url@samestyle\endcsname
\providecommand{\newblock}{\relax}
\providecommand{\bibinfo}[2]{#2}
\providecommand{\BIBentrySTDinterwordspacing}{\spaceskip=0pt\relax}
\providecommand{\BIBentryALTinterwordstretchfactor}{4}
\providecommand{\BIBentryALTinterwordspacing}{\spaceskip=\fontdimen2\font plus
\BIBentryALTinterwordstretchfactor\fontdimen3\font minus
  \fontdimen4\font\relax}
\providecommand{\BIBforeignlanguage}[2]{{%
\expandafter\ifx\csname l@#1\endcsname\relax
\typeout{** WARNING: IEEEtran.bst: No hyphenation pattern has been}%
\typeout{** loaded for the language `#1'. Using the pattern for}%
\typeout{** the default language instead.}%
\else
\language=\csname l@#1\endcsname
\fi
#2}}
\providecommand{\BIBdecl}{\relax}
\BIBdecl

\bibitem{kunal2019align}
K.~Kunal, M.~Madhusudan, A.~K. Sharma, W.~Xu, S.~M. Burns, R.~Harjani, J.~Hu,
  D.~A. Kirkpatrick, and S.~S. Sapatnekar, ``Align: Open-source analog layout
  automation from the ground up,'' in \emph{DAC}, 2019, pp. 1--4.

\bibitem{crossley2013bag}
J.~Crossley, A.~Puggelli, H.-P. Le, B.~Yang, R.~Nancollas, K.~Jung, L.~Kong,
  N.~Narevsky, Y.~Lu, N.~Sutardja \emph{et~al.}, ``Bag: A designer-oriented
  integrated framework for the development of ams circuit generators,'' in
  \emph{ICCAD}.\hskip 1em plus 0.5em minus 0.4em\relax IEEE, 2013, pp. 74--81.

\bibitem{chang2018bag2}
E.~Chang, J.~Han, W.~Bae, Z.~Wang, N.~Narevsky, B.~NikoliC, and E.~Alon,
  ``Bag2: A process-portable framework for generator-based ams circuit
  design,'' in \emph{IEEE Custom Integrated Circuits Conference (CICC)}.\hskip
  1em plus 0.5em minus 0.4em\relax IEEE, 2018, pp. 1--8.

\bibitem{xu2019magical}
H.~Chen, M.~Liu, X.~Tang, K.~Zhu, A.~Mukherjee, N.~Sun, and D.~Z. Pan,
  ``Magical 1.0: An open-source fully-automated ams layout synthesis framework
  verified with a 40-nm 1gs/s $\delta\sigma$ adc,'' in \emph{IEEE Custom
  Integrated Circuits Conference (CICC)}, 2021, pp. 1--2.

\bibitem{liu2021parasitic}
M.~Liu, W.~J. Turner, G.~F. Kokai, B.~Khailany, D.~Z. Pan, and H.~Ren,
  ``Parasitic-aware analog circuit sizing with graph neural networks and
  bayesian optimization,'' in \emph{DATE}.\hskip 1em plus 0.5em minus
  0.4em\relax IEEE, 2021, pp. 1372--1377.

\bibitem{ranjan2004fast}
M.~Ranjan, W.~Verhaegen, A.~Agarwal, H.~Sampath, R.~Vemuri, and G.~Gielen,
  ``Fast, layout-inclusive analog circuit synthesis using pre-compiled
  parasitic-aware symbolic performance models,'' in \emph{DATE}, vol.~1.\hskip
  1em plus 0.5em minus 0.4em\relax IEEE, 2004, pp. 604--609.

\bibitem{hakhamaneshi2019bagnet}
K.~Hakhamaneshi, N.~Werblun, P.~Abbeel, and V.~Stojanovi{\'c}, ``Bagnet:
  Berkeley analog generator with layout optimizer boosted with deep neural
  networks,'' in \emph{ICCAD}.\hskip 1em plus 0.5em minus 0.4em\relax IEEE,
  2019, pp. 1--8.

\bibitem{settaluri2020autockt}
K.~Settaluri, A.~Haj-Ali, Q.~Huang, K.~Hakhamaneshi, and B.~Nikolic, ``Autockt:
  deep reinforcement learning of analog circuit designs,'' in
  \emph{DATE}.\hskip 1em plus 0.5em minus 0.4em\relax IEEE, 2020, pp. 490--495.

\bibitem{bengio2017deep}
Y.~Bengio, I.~Goodfellow, and A.~Courville, \emph{Deep learning}.\hskip 1em
  plus 0.5em minus 0.4em\relax MIT press Cambridge, MA, USA, 2017, vol.~1.

\bibitem{chai2022circuitnet}
Z.~Chai, Y.~Zhao, Y.~Lin, W.~Liu, R.~Wang, and R.~Huang, ``Circuitnet: An
  open-source dataset for machine learning applications in electronic design
  automation (eda),'' \emph{SCIENCE CHINA Information Sciences}, vol.~65,
  no.~12, 2022.

\bibitem{boyd2001optimal}
S.~P. Boyd, T.~H. Lee \emph{et~al.}, ``Optimal design of a cmos op-amp via
  geometric programming,'' \emph{IEEE TCAD}, vol.~20, no.~1, pp. 1--21, 2001.

\bibitem{wang2020gcn}
H.~Wang, K.~Wang, J.~Yang, L.~Shen, N.~Sun, H.-S. Lee, and S.~Han, ``Gcn-rl
  circuit designer: Transferable transistor sizing with graph neural networks
  and reinforcement learning,'' in \emph{DAC}.\hskip 1em plus 0.5em minus
  0.4em\relax IEEE, 2020, pp. 1--6.

\bibitem{li2021circuit}
Y.~Li, Y.~Lin, M.~Madhusudan, A.~Sharma, S.~Sapatnekar, R.~Harjani, and J.~Hu,
  ``A circuit attention network-based actor-critic learning approach to robust
  analog transistor sizing,'' in \emph{ACM/IEEE Workshop on Machine Learning
  for CAD (MLCAD)}.\hskip 1em plus 0.5em minus 0.4em\relax IEEE, 2021, pp.
  1--6.

\bibitem{DAC_2019}
S.~Zhang, W.~Lyu, F.~Yang, C.~Yan, D.~Zhou, X.~Zeng, and X.~Hu, ``An efficient
  multi-fidelity bayesian optimization approach for analog circuit synthesis,''
  in \emph{DAC}, 2019, pp. 1--6.

\bibitem{zhang2020efficient}
S.~Zhang, F.~Yang, D.~Zhou, and X.~Zeng, ``An efficient asynchronous batch
  bayesian optimization approach for analog circuit synthesis,'' in
  \emph{DAC}.\hskip 1em plus 0.5em minus 0.4em\relax IEEE, 2020, pp. 1--6.

\bibitem{bayes_fusion}
F.~Wang, W.~Zhang, S.~Sun, X.~Li, and C.~Gu, ``Bayesian model fusion:
  Large-scale performance modeling of analog and mixed-signal circuits by
  reusing early-stage data,'' in \emph{DAC}, 2013, pp. 1--6.

\bibitem{transfer_bayesian_sampling}
J.~Liu, M.~Hassanpourghadi, Q.~Zhang, S.~Su, and M.~S.-W. Chen, ``Transfer
  learning with bayesian optimization-aided sampling for efficient ams circuit
  modeling,'' in \emph{ICCAD}, 2020, pp. 1--9.

\bibitem{bergstra2011algorithms}
J.~Bergstra, R.~Bardenet, Y.~Bengio, and B.~K{\'e}gl, ``Algorithms for
  hyper-parameter optimization,'' \emph{Advances in neural information
  processing systems}, vol.~24, 2011.

\bibitem{snoek2012practical}
J.~Snoek, H.~Larochelle, and R.~P. Adams, ``Practical bayesian optimization of
  machine learning algorithms,'' \emph{Advances in neural information
  processing systems}, vol.~25, 2012.

\bibitem{shahriari2015taking}
B.~Shahriari, K.~Swersky, Z.~Wang, R.~P. Adams, and N.~De~Freitas, ``Taking the
  human out of the loop: A review of bayesian optimization,'' \emph{Proceedings
  of the IEEE}, vol. 104, no.~1, pp. 148--175, 2015.

\bibitem{zoopt}
Y.-R. Liu, Y.-Q. Hu, H.~Qian, C.~Qian, and Y.~Yu, ``Zoopt: Toolbox for
  derivative-free optimization,'' \emph{SCIENCE CHINA Information Sciences},
  vol.~65, no.~10, p. 207101, 2022.

\bibitem{rasmussen2003gaussian}
C.~E. Rasmussen, ``Gaussian processes in machine learning,'' in \emph{Summer
  school on machine learning}.\hskip 1em plus 0.5em minus 0.4em\relax Springer,
  2003, pp. 63--71.

\bibitem{mockus1978application}
J.~Mockus, V.~Tiesis, and A.~Zilinskas, ``The application of bayesian methods
  for seeking the extremum,'' \emph{Towards global optimization}, vol.~2, no.
  117-129, p.~2, 1978.

\bibitem{kushner1964new}
H.~J. Kushner, ``A new method of locating the maximum point of an arbitrary
  multipeak curve in the presence of noise,'' 1964.

\bibitem{thompson1933likelihood}
W.~R. Thompson, ``On the likelihood that one unknown probability exceeds
  another in view of the evidence of two samples,'' \emph{Biometrika}, vol.~25,
  no. 3/4, pp. 285--294, 1933.

\bibitem{hennig2012entropy}
P.~Hennig and C.~J. Schuler, ``Entropy search for information-efficient global
  optimization.'' \emph{Journal of Machine Learning Research}, vol.~13, no.~6,
  2012.

\bibitem{kingma2013auto}
D.~P. Kingma and M.~Welling, ``Auto-encoding variational bayes,'' \emph{arXiv
  preprint arXiv:1312.6114}, 2013.

\bibitem{rezende2014stochastic}
D.~J. Rezende, S.~Mohamed, and D.~Wierstra, ``Stochastic backpropagation and
  approximate inference in deep generative models,'' in \emph{International
  conference on machine learning}.\hskip 1em plus 0.5em minus 0.4em\relax PMLR,
  2014, pp. 1278--1286.

\bibitem{srivastava1973estimation}
R.~Srivastava, ``Estimation of probability density function based on random
  number of observations with applications,'' \emph{International Statistical
  Review/Revue Internationale de Statistique}, pp. 77--86, 1973.

\bibitem{zhu2020effective}
K.~Zhu, H.~Chen, M.~Liu, X.~Tang, N.~Sun, and D.~Z. Pan, ``Effective
  analog/mixed-signal circuit placement considering system signal flow,'' in
  \emph{ICCAD}.\hskip 1em plus 0.5em minus 0.4em\relax IEEE, 2020.

\bibitem{chen2020toward}
H.~Chen, K.~Zhu, M.~Liu, X.~Tang, N.~Sun, and D.~Z. Pan, ``Toward
  silicon-proven detailed routing for analog and mixed-signal circuits,'' in
  \emph{ICCAD}.\hskip 1em plus 0.5em minus 0.4em\relax IEEE, 2020.

\bibitem{liu2020s}
M.~Liu, W.~Li, K.~Zhu, B.~Xu, Y.~Lin, L.~Shen, X.~Tang, N.~Sun, and D.~Z. Pan,
  ``{S3DET}: Detecting system symmetry constraints for analog circuits with
  graph similarity,'' in \emph{ASPDAC}.\hskip 1em plus 0.5em minus 0.4em\relax
  IEEE, 2020, pp. 193--198.

\end{thebibliography}
